\documentstyle[prl,aps,epsf,floats]{revtex}
\begin{document}
\draft
\twocolumn[\hsize\textwidth\columnwidth\hsize\csname @twocolumnfalse\endcsname
\title{Conjectures for the microscopic theory of   
       high temperature     \\
       superconductivity}
\author{Bumsoo Kyung}
\address{D\'{e}partement de physique and Centre de recherche 
sur les propri\'{e}t\'{e}s \'{e}lectroniques \\
de mat\'{e}riaux avanc\'{e}s. 
Universit\'{e} de Sherbrooke, Sherbrooke, Qu\'{e}bec, Canada J1K 2R1}
\date{August 28, 2000}
\maketitle
\begin{abstract}

    Based on experimental results and our previous theoretical work, 
a microscopic theory of high temperature superconductivity is conjectured.
In this conjecture,
superconducting and antiferromagnetic long-range orders are driven by 
interlayer coupling.
Strictly in two dimensions, the microscopic Hubbard model has an 
(resonating valence bond) insulator-to-metal transition 
at $x=x_{c}$ near optimal doping 
for zero temperature, leading to a quantum critical 
point, and one of the crossover lines is given by the pseudogap 
temperature $T^{*}$.
We argue that various singular and non-Fermi liquid properties observed  
near optimal doping are due to the presence of this quantum critical point.
In our conjecture,
the crossover line $T^{*}$ also practically divides the superconducting  
region into two,
depending on the doping level with respect to $x_{c}$.
For $x \leq x_{c}$ the superconducting state has significant  
antiferromagnetic correlations, while 
for $x > x_{c}$ it has virtually no antiferromagnetic correlations, thus 
justifying the conventional 
BCS theory based on the noninteracting electrons.
Inelastic neutron scattering resonance and systematically reduced 
superfluid density in the superconducting state below $x_{c}$ 
have their natural explanations in the 
present scenario.
The present approach supports
interlayer pair tunneling model where  
the superconducting condensation energy comes from the lowering of the 
c-axis kinetic energy in the superconducting state.
Comparison of the present scenario with some of the leading theories 
based on the Hubbard and $t-J$ models is given. 
The generic features of
both hole-doped and electron-doped cuprates as well as 
heavy-fermion superconductors may be understood in the 
{\em unified} framework within the present picture.
\end{abstract}
\vskip2pc]
\narrowtext
\section{Introduction}
\label{section1}

   Since the discovery of high temperature superconductivity in copper 
oxide compounds,\cite{Bednorz:1986}  
enormous experimental and theoretical effort has been  
made in order to understand 
various anomalous behaviors in normal and superconducting (SC) 
states of these materials. 
Right after its discovery,
it was recognized that the high temperature cuprate
superconductors have in common layered perovskite-like crystal 
structures which consist of conducting CuO$_{2}$ planes separated by 
layers of other elements.
These latter layers function as charge reservoirs and mobile 
charge carriers (holes or electrons) supplied from them 
are believed to reside mainly 
within the CuO$_{2}$ planes.
Through extensive worldwide effort as well as the improvement of 
sample quality over the years,
many consensuses 
have been reached in experimental side.
First, let us start by showing a generic phase diagram 
(Fig. ~\ref{fig1}(a))
of a hole-doped cuprate Ln$_{2-x}$Sr$_{x}$CuO$_{4}$ in the 
doping ($x=1-n$) and temperature ($T$) plane.\cite{Almasan:1991}
Since this compound has only one CuO$_{2}$ layer in a unit cell,  
it is one of the best materials to study physics taking place
in a two-dimensional CuO$_{2}$ plane.
In other hole-doped compounds, the presence of multi-layers and 
CuO chains makes it more complicated to extract the intrinsic features
associated with a single CuO$_{2}$ plane.
But the CuO$_{2}$ plane in Ln$_{2-x}$Sr$_{x}$CuO$_{4}$
is still not totally isolated from other CuO$_{2}$ planes
in different unit cells.

   Near half-filling and at low temperature, antiferromagnetic (AF) long-range
order appears with 
$T_{N}=$ 250-300 K at $x=0$.
It is destroyed by $2 \%$ doping concentration.
When $x$ reaches $0.06$, SC long-range order starts to 
appear, and it is also destroyed by $30 \%$ doping. 
In between them, $T_{c}$ reaches a maximum value of 40 K at $x \simeq 0.16$.
The SC gap was found to have mainly 
$d$-wave character with possibility
of a small mixture of other angular momentum
states,\cite{Levi:1993,%
Shen:1993,Marshall:1996}
in contrast to conventional BCS superconductors\cite{Bardeen:1957} 
with an isotropic
$s$-wave gap.
The low temperature phase between $T_{N}$ and $T_{c}$ is often 
designated as spin glass (SG) phase.

   Various recent experiments also show the existence of a crossover 
temperature $T^{*}$ larger than $T_{c}$ in a doping range of
$x=0$ to $x \simeq 0.18-0.19 $.%
\cite{Ding:1996,Loeser:1996,%
Loram:1993,%
Renner:1998,%
Takigawa:1991,%
Homes:1993}
Below this pseudogap temperature $T^{*}$,
the low frequency spectral weight begins to
be strongly suppressed.
Surprisingly the doping dependences of $T^{*}$ and $T_{c}$ are
completely different in spite of 
their close relationship suggested by angle resolved photoemission 
(ARPES)\cite{Ding:1996,Loeser:1996},
tunneling\cite{Renner:1998} and NMR experiments.\cite{Takigawa:1991}
At optimal doping where $T_{c}$ is maximum, various non-Fermi 
liquid (NFL) properties are observed in the normal state.
These include the linear temperature dependence of ab-plane resistivity,
the quadratic $T$ dependence of Hall angle and so on
up to 1000 K.
Far beyond optimal doping, the normal state properties are well 
described by the conventional Landau Fermi liquid.
Several anomalous behaviors have been observed in the SC state as well.
Near optimal doping and at underdoping, 
a sharp resonance (almost energy resolution limited) is 
observed in inelastic neutron scattering
experiments.\cite{Mook:1993} 
Furthermore in the same doping range, the superfluid density $n_{s}$ is 
systematically suppressed with decreasing doping in spite of 
increasing SC gap amplitude.
In the overdoped regime, however, the SC properties appear to be well explained 
by the conventional weak coupling BCS theory.

   Although they are not usually addressed in the context of a
phase diagram, the following experiments deserve special attention in order 
to understand the complete picture of the physics.
A recent ARPES experiment for an insulating
cuprate
Sr$_{2}$CuO$_{2}$Cl$_{2}$\cite{Wells:1995}
showed nearly isotropic and quite similar band dispersions
along $(\pi/2,\pi/2)-(\pi,0)$ and
$(\pi/2,\pi/2)-(0,0)$ directions.
Furthermore
a $d$-wave-like modulation of the insulating gap
in Ca$_{2}$CuO$_{2}$Cl$_{2}$\cite{Ronning:1998}
requires even at half-filling the presence of strong pairing fluctuations.
In another important experiment,
a 61 Tesla pulsed magnetic field suppressed superconductivity in 
Ln$_{2-x}$Sr$_{x}$CuO$_{4}$ single crystals and revealed
an insulator-to-metal (IM) crossover for both ab-plane 
resistivity $\rho_{ab}$ and c-axis resistivity $\rho_{c}$
near optimal doping.\cite{Boebinger:1996}
The latter experiment may shed some insight  
into the underlying physics of normal state 
cuprates which is usually masked by the presence of SC long-range order.
Any successful theory should explain all these features in a natural 
and yet unified way.

   As to the theoretical side, 
right after the discovery of high temperature superconductors,
Anderson\cite{Anderson:1987} first proposed 
the one-band Hubbard model
as the simplest Hamiltonian which might capture the correct
low energy physics of copper oxides.
Zhang and Rice\cite{Zhang:1988}
also derived an effective $t-J$ Hamiltonian from
the more realistic three-band Hubbard model.
The $t-J$ Hamiltonian was already known to be the large $U$ limit
of the Hubbard Hamiltonian under certain assumptions.
The Hubbard model is described by the Hamiltonian
in which $c_{i,\sigma}$ destroys an electron at site $i$ with spin $\sigma$
on a two-dimensional square lattice
\begin{eqnarray}
 H = -t\sum_{\langle i,j \rangle,\sigma}c^{+}_{i,\sigma}
      c_{j,\sigma}
    +U\sum_{i}
     c^{+}_{i,\uparrow}
         c_{i,\uparrow}
     c^{+}_{i,\downarrow}
         c_{i,\downarrow} \; .
                                                           \label{eq10}
\end{eqnarray}
$t$ is a hopping parameter between nearest neighbors $\langle i,j \rangle$ 
and $U$ denotes local Coulomb repulsion.
It is believed that a realistic strength of the Coulomb repulsion
lies in between the weak and strong coupling regimes, namely,
$U \sim W-2W$ where $W$ is the bandwidth of $8t$ in two dimensions.
So far many theories have been proposed to understand the various
anomalous properties of the cuprates by directly invoking 
the Hubbard model or its variants   
(one-band and three-band Hubbard, and $t-J$ models),  
or other phenomenological models.
However, at present there is no consensus on which kind of approach is most 
appropriate for a given Hamiltonian or even 
on which Hamiltonian is most 
relevant for the cuprates. 
At the end of this paper,
comparison of our approach with some leading 
theories
based on the Hubbard and $t-J$ models
will be given.
\section{Review of our previous work}
\label{section2}

    First, let us begin by 
summarizing the main results of our previous work\cite{Kyung:2000-2} 
and presenting a proposed phase diagram 
in order to set the stage for the present study.
As a first step to the microscopic theory of high-$T_{c}$ superconductivity,
it is of great importance to identify pairing    
interaction which may eventually lead to a SC instability at low 
temperature,
just like the 
Cooper problem in the development of the BCS theory.
In a recent study, 
we examined the possibility of extracting pairing interaction
{\em directly} from the Coulomb repulsion itself without the exchange of
bosonic degrees of freedom such as spin fluctuations.

   We found that
among several gap symmetries
pairing interaction with 
the $d_{x^2-y^2}$ ($\phi_{d}(\vec{k})=\cos k_{x} - \cos k_{y}$) symmetry
is most strongly induced from the local Coulomb repulsion $U$.
It is consistent with a weak coupling renormalization 
group (RG) calculation.\cite{Schulz:1987}
Due to its direct Coulombic origin,
this mechanism for pairing correlations
is similar to the lattice version of the Kohn-Luttinger
theorem\cite{Kohn:1965}. 
However,
the details are somewhat different. The Kohn-Luttinger
theory requires a sharp Fermi surface and a resulting  
long-range oscillatory behavior in real space of the 
effective interaction, while our approach does not.

   AF correlations are also induced 
from the local Coulomb repulsion
{\em at the same time} as $d$-wave pairing correlations are.
Both the induced pairing and AF correlations are found to 
{\em increase with decreasing doping}.
The simultaneous induction of
pairing and AF correlations from
the same Coulomb repulsion may be understood in the context of
the $t-J$ model.
Projecting out the doubly occupied sites in the large $U$ limit of
the Hubbard model generates or {\em induces} the Heisenberg term
in the $t-J$ model. It has in general both pairing and AF correlations.
The increasing pairing and AF correlations with
decreasing doping found in our previous work
is also realized in the $t-J$ model (in a relative sense) as
decreasing hopping probability with decreasing doping.
In that paper, we argued that
the induced pairing correlations are the microscopic
origin of pseudogap behavior in underdoped cuprates.
This is because
the pairing fluctuation origin of the pseudogap and its doping dependence
are consistent with many experiments.
This is also because of the excellent agreement of our calculations with
the observed energy dispersion of the insulating cuprates,
Sr$_{2}$CuO$_{2}$Cl$_{2}$\cite{Wells:1995} and
Ca$_{2}$CuO$_{2}$Cl$_{2}$.\cite{Ronning:1998}
It is expected that the high energy pseudogap 
at half-filling continuously evolves into a relatively smaller 
pseudogap away from half-filling.

   Based on the doping dependences of the induced pairing and
AF correlations,
a phase diagram was proposed for the high temperature superconductors,
which is shown in Fig. ~\ref{fig1}(b).
First, far away from half-filling the short-range correlations due to
local Coulomb repulsion are very weak and don't produce
strong enough pairing interaction so that $T_{c}$ vanishes.
With decreasing doping, they start to induce both pairing
and AF fluctuations, but the latter is probably not
strong enough to compete with the former. Thus up to
optimal doping level, local Coulomb repulsion plays as a driving force to
superconductivity and $T_{c}$ keeps growing.
When electrons enter into the underdoping regime,
the induced AF fluctuations start to play as a SC phase coherence-breaker
by creating {\em locally} the spin density wave (SDW) state.  
It strongly breaks time-reversal symmetry and 
causes $T_{c}$ to decrease.
As far as $T^{*}$ is concerned,
beyond optimal doping,
induced AF fluctuations are weak and thus
the pseudogap temperature $T^{*}$ closely follows
$T_{c}$.
The strength of induced pairing interaction increases
with decreasing doping.
This makes $T^{*}$ increase, creating a huge pseudogap region
between $T^{*}$ and $T_{c}$.

   Near half-filling and at low temperature,
AF long-range order is stabilized by the 
three-dimensional effect.
The AF spin fluctuations become strong more or less
around the AF phase boundary, leading to another kind of a crossover
temperature $T_{RVB}$.\cite{Comment5}
Below $T_{RVB}$ 
both AF and pairing fluctuations are strong
and coexist.
It is expected that 
the SC long-range order completely vanishes,
when it enters into the RVB-like region.
The $T=0$ ground state between the AF and SC phase boundaries 
(in a clean sample)
is the RVB insulator in which an insulating gap is identified with the
spingap $\Delta$. 
Note that $T_{c}$, $T^{*}$, $T_{N}$ and $T_{RVB}$ are determined as a 
result of close competition between the induced pairing and AF correlations
and that any imbalance of the two correlations may 
drastically change the phase diagram.
The pseudogap behavior in this scenario is as generic as 
AF correlations, because they are {\it simultaneously} induced 
from the same local  
Coulomb repulsion.
In our picture
$T^{*}$ is different from a temperature below which
the critical behavior (O(2) symmetry)
of SC long-range order\cite{Corson:1999} is found.
In this previous work, however, interlayer coupling effect and the possibility 
of existence of a quantum critical point (QCP) were not considered.
\section{RVB insulator-to-metal transition near optimal doping
         at zero temperature: QCP in the 
         two-dimensional Hubbard model}
\label{section3}

   In principle,
the above phase diagram 
resulting from the close competition between 
the induced pairing and AF correlations
should be {\em quantitatively} similar for different cuprates
with similar $U$'s. 
This is because these correlations are
completely determined by the same Coulomb repulsion $U$ at given $x$ and $T$.
For most of the 
hole-doped cuprates,
$J=4t^{2}/U$ lies in the range of 120-140 meV, indicating quite 
similar $U$'s with $20 \%$ variation at most.
One the other hand, $T_{c}$'s, which are expected to be similar 
with $20 \%$ difference, widely vary from 40 K for 
Ln$_{2-x}$Sr$_{x}$CuO$_{4}$ to 133 K for 
HgBa$_{2}$Ca$_{2}$Cu$_{3}$O$_{8}$.
This particular feature strongly suggests that 
the true SC long-range 
order is driven by interlayer coupling.
It is consistent with the general trend that  
the cuprate compounds with more CuO$_{2}$ planes in a unit cell
show higher $T_{c}$.
In the present paper, the interlayer coupling means not only 
the coupling between CuO$_{2}$ planes in a unit cell, but also
between other CuO$_{2}$ planes in different unit cells.
Strictly in two dimensions,
AF long-range order cannot appear at finite temperature  
due to the 
Mermin-Wagner theorem\cite{Mermin:1966}, while 
SC long-range order can be stabilized at finite temperature by
the Kosterlitz-Thouless phase transition.\cite{Kosterlitz:1973}

   However, there is possibility that
SC long-range order is also destroyed at finite
temperature mainly by the
phase coherence-breaking effect of
the induced AF fluctuations and partly by
quantum fluctuations.
This is consistent with the rigorous result  
by Su and Suzuki.\cite{Su:1998} 
These authors proved the nonexistence of
d$_{x^{2}-y^{2}}$ superconductivity (long-range order)  
at any nonzero temperature in 
the two-dimensional Hubbard model.
Note that the proof by Su and Suzuki does not exclude increasing 
pairing correlations\cite{Comment6.5} with decreasing temperature  
just like the AF 
correlations for a half-filled Hubbard band.
There is also possibility that strictly in two dimensions
both SC and AF long range orders are absent even at
zero temperature.
This is because AF fluctuatings with spin triplet character
(pairing fluctuatings with spin singlet)
destroy SC long-range coherence with spin singlet character
(AF long-range coherence with spin triplet).
In a sense, the induced pairing and AF correlations
may act in two dimensions   
like frustrating the long-range coherence of the other correlations.
Unlike the SC long-range order,
the absence of AF long-range order at zero temperature
is not so crucial for our argument.
Thus in our first conjecture, 
only by the three-dimensional effect
are stabilized 
the AF and SC long-range orders 
at finite temperature and even at zero temperature. 

   Now in order to understand the intrinsic electronic properties in  
a totally isolated 
CuO$_{2}$ plane, let us do a {\em thought experiment} on the phase diagram 
(Fig. ~\ref{fig1}(a)) by slowly turning off the interlayer coupling.
It is clear that 
slow turning off of the interlayer coupling reduces  
the AF and SC phase boundaries. 
At the same time the region of the RVB insulating ground 
state expands, as shown in Fig. ~\ref{fig2}(a).
When the interlayer coupling is completely disconnected from neighboring 
CuO$_{2}$ planes in the same unit cell or in other unit cells, 
the AF and SC long-range orders totally disappear.
And the RVB insulating ground state persists from half-filling all the 
way up to $x=x_{c}$ near optimal doping, as shown in Fig. ~\ref{fig2}(b).
Considering $T^{*}$ meets $T_{c}$ at slightly overdoping region
and its curvature,
$x=x_{c}$ should be somewhat larger than 
optimal doping ($x_{c} \simeq 0.18-0.19$).
Since $T^{*}$ involves a short distance scale or rather a high energy scale,
$T^{*}$ obtained in two dimensions is not expected to change 
substantially by a small interlayer coupling. 

   Thus in our second conjecture,
the two-dimensional Hubbard model has an RVB insulator-to-metal
transition
at $x=x_{c}$ for $T=0$, leading to a QCP,
and one of the crossover lines is given by the pseudogap temperature 
$T^{*}$.
This feature is presumably what Boebinger {\em et al.}\cite{Boebinger:1996}  
found in their experiment of an
insulator-to-metal crossover near optimal doping.
As is well known from an extensive study of QCP   
by Sachdev,\cite{Sachdev:1999}  
an immediate consequence from 
the presence of QCP is to create two crossover lines which occur 
at $T \sim |\Delta|^{z\nu}$, where $z$ is the dynamical exponent and 
$\nu$ is the correlation length exponent (Fig. ~\ref{fig2}(c)).
The crossover line above the RVB insulating ground state 
is identified with the pseudogap temperature $T^{*}$.
The state above the QCP, sandwiched by the two crossover lines,
is described by placing the $\Delta=0$ scale-invariant critical theory at 
nonzero temperature.
This is probably the region where 
various singular\cite{Aeppli:1997} and NFL properties have been observed
in the normal state and phenomenological marginal Fermi liquid (MFL)
theory by Varma {\em et al.}\cite{Varma:1989}  
works well.
The fact that 
the pseudogap temperature $T^{*}$ naturally divides the pseudogap region
and the NFL region, justifies  
a commonly used experimental method\cite{Timusk:1999} 
of determining the pseudogap temperature.
For instance, in dc resistivity measurements, $T^{*}$ is a temperature
where the resistivity starts to deviate from a linear temperature  
dependence.

   Since we have identified strictly in two dimensions the QCP
of the RVB insulator-to-metal and 
concomitant NFL behaviors in optimally doped cuprates,
let us turn on the interlayer coupling 
in order to find the realistic phase diagram
in quasi-two dimensions.
Before doing that, it is important to point out   
the differences in electronic properties  
below and above $x_{c}$ at low temperature. 
In our third conjecture, at low temperature ($T^{*} \simeq 0$),
the crossover line $T^{*}$ roughly divides the doping-temperature
plane into two regions,
one with AF correlations and the other without them.
The identification of $T^{*}$ at low temperature as a crossover
of the AF correlations
may be understood by the following two observations.

   In our previous work,\cite{Kyung:2000-2}
another crossover line $T_{RVB}$, which is not specifically addressed 
in this paper, was defined as a line where  
the AF correlations are just strong enough to destroy
the SC long-range phase coherence,
as shown in Fig. ~\ref{fig1}(b).
When the interlayer coupling is completely turned off, it is easy
to see $T_{RVB}$ approach $T^{*}$ at low temperature.\cite{Comment9}
As another supporting evidence, we\cite{Kyung:2000-3} and 
other groups\cite{Inui:1988,Inaba:1996} found that most of the low temperature
region including the SC state below $x \simeq 0.2$ is inside of the AF phase
in a mean-field study of the $t-J$ model.
In a mean-field approximation, long-range order
already sets in when
the corresponding correlation length reaches roughly one
lattice spacing.
This dictates the above mean-field result to be interpreted as 
the presence of short-range AF correlations for $x \leq 0.2$
at low temperature.
If this conjecture turns out to be correct,
there is possibility that at low temperature and  
near $x \simeq x_{c}$ a band Kondo effect is 
realized in finite dimensions, as is done in infinite 
dimensions.\cite{Georges:1996}
In the band Kondo effect, a nonvanishing local magnetic moment at site $i$,
$\langle S^{z}_{i} \rangle$, 
due to local Coulomb repulsion is effectively screened out by
surrounding (but the same kind of electrons at site $i$) itinerant electrons.

   From a purely pairing correlation point of view,
the pairing correlations become singular with decreasing temperature
at any doping concentration.
This makes
the crossover region of the pairing correlations much wider
than that of the AF correlations
at low temperature near $x \simeq x_{c}$.
As another supporting evidence on this point,
we again invoke our previous mean-field phase diagram.\cite{Kyung:2000-3}
The mean-field $T_{c}$ line, which is interpreted as a crossover of
short-range pairing correlations,
extends up to $x \simeq 0.35$ beyond $x=x_{c}$ where
the mean-field AF order vanishes.
It makes the crossover region of pairing correlations broad
with respect to $x=x_{c}$ at low temperature.

   In this situation, a potential location with the highest $T_{c}$, when 
the interlayer coupling turned on, is near $x=x_{c}$ in which
the phase coherence-breaking AF correlations nearly vanish but 
the pairing correlations are still robust.
The resulting phase diagram (Fig. ~\ref{fig1}(a)) 
will look like one where $T^{*}$ falls from 
a high value onto the $T_{c}$ line rather than the other where 
$T^{*}$ smoothly merges with $T_{c}$ in the slightly overdoped region, as 
recently argued by Tallon and Loram.\cite{Tallon:2000}
In our scenario $T_{c}$ is never part of the $T^{*}$ line.
For different compounds, different strength of interlayer pairing hopping
drives SC long-ranger order
in the {\it same} background of the two-dimensional electron system.
This leads to
a universal relation\cite{Zhang:1993}
$T_{c}/T^{\mbox{max}}_{c} = 1-82.6(x-0.16)^{2}$, when $T_{c}$ is
scaled by $T^{\mbox{max}}_{c}$.
Away from half-filling,
the AF correlations manifest their existence most strongly
in the SC state, because
they easily destroy the SC long-range phase coherence.
Our scenario also predicts that due to scatterings with AF fluctuations,
quasiparticle scattering rate remains finite for $x \leq x_{c}$
even in a clean sample and at $T=0$
and it is universal
(up to the variation of $J$ or $U$ for different compounds).
For $x \leq x_{c}$, in our picture,
the Landau quasiparticle with nonvanishing quasiparticle residue
does not exist in the normal state
due to strong scatterings with pairing and AF fluctuations or
strong influence from the QCP.
But it can be stabilized in the SC state ($T < T_{c}$) where
its coherence is restored through a pair hopping process along the c-axis.
Because of the strong influence from the QCP,
the NFL takes a special structure near optimal doping, presumably
the marginal Fermi liquid by Varma {\em et al.}\cite{Varma:1989}
\section{Inelastic neutron scattering resonance and reduced 
         superfluid density in the SC state}
\label{section4}

   As long as the strength of pairing correlations is concerned, it lies in 
the intermediate to weak coupling regimes near optimal doping,
as shown in Ref. ~\cite{Kyung:2000-2}.
If a screening effect is correctly taken into account in obtaining  
$V_{ind}$ (by considering particle-particle and particle-hole 
channels on equal footing), $V_{ind}$ can be further reduced 
near and beyond optimal doping.
Thus the crucial point in proper understanding of the SC state is 
not the validity of the weak coupling BCS theory, but 
more importantly whether Cooper pairs are constructed from 
antiferromagnetically correlated electrons or not.
In fact 
depending on the doping concentration with respect 
to $x=x_{c}$,
the SC state can be qualitatively different.
For $x \leq x_{c}$ the SC state has significant AF correlations, while 
for $x > x_{c}$ it has virtually no AF correlations, thus 
justifying the conventional 
BCS theory based on the noninteracting electrons.

   First, we discuss a strong inelastic  
neutron scattering resonance\cite{Mook:1993} 
observed in the SC state of the underdoped and optimally doped 
cuprates ($x \leq x_{c}$).
The striking feature of a sharp (almost energy resolution limited) 
resonance peak\cite{Bourges:2000} is that its energy $E_{r}$ does not 
shift towards lower energy when approaching $T_{c}$, but its intensity
vanishes upon heating at $T_{c}$.
On experimental grounds, the resonance peak is often considered as a 
collective mode in the SC state.
The inelastic neutron scattering experiment measures the 
dynamical spin-spin correlation function
\begin{eqnarray}
  S(\vec{q},\omega) = \sum_{n}
                 | \langle n | S^{+}_{\vec{q}} | 0 \rangle |^{2}
                 \delta(\omega-\omega_{n0})
                           \; ,
                                                           \label{eq20}
\end{eqnarray}
where $| 0 \rangle$ and $| n \rangle$ are the ground and excited states 
of the system, and 
$\omega_{n0}$ is the energy difference between these states, and 
$S^{+}_{\vec{q}} = \sum_{\vec{k}} c^{+}_{\vec{k}+\vec{q},\uparrow}
                                  c    _{\vec{k},\downarrow}$.

   In the present scenario,
the neutron scattering resonance 
is caused by the presence of AF correlations in the 
SC state for $x \leq x_{c}$.
In a recent mean-field study of the interplay between antiferromagnetism
and $d$-wave superconductivity,\cite{Kyung:2000-1} we noted that
when $d$-wave SC order and AF order coexist\cite{Comment10},
another order parameter with
a spin-triplet,
$\langle c_{\vec{k}+\vec{Q},\uparrow}c_{-\vec{k},\downarrow} \rangle $,
is dynamically generated.
$\vec{Q}$ is the AF wave vector $(\pi,\pi)$ in two dimensions.
This is a generic feature for fermionic systems and cannot be obtained in
a purely bosonic description of pairing and AF correlations.
As long as the two correlations are strong (but both of
them need not be in long-range ordered states),
the spin-triplet correlations can be robust.
Since both pairing and AF correlations are present below $x_{c}$ as noted
in the previous Section, in general the ground state has three different
(pairing, AF and spin-triplet) correlations.
To have a maximum transition amplitude
$| \langle n | S^{+}_{\vec{q}=\vec{Q}} | 0 \rangle |$,
one finds two possible ways of transition between the three correlations,
namely, AF $\leftrightarrow$ AF and pairing $\leftrightarrow$ spin-triplet.
The former is most effective near half-filling, due to the presence of
the AF long-range order in that region.
On the other hand, the latter is most operative near $x=x_{c}$,
because the effective strength of SC long-range order 
is strongest there.\cite{Tallon:2000}
Between at half-filling and at $x=x_{c}$,
the two ways of transition compete.

   Near $x=x_{c}$ where a strong neutron resonance was observed
and the second way of transition is most effective,
the resonance energy is given by twice of the maximum gap energy,
and its intensity by the product of $d$-wave SC order
and SDW order parameters\cite{Kyung:2000-1}.
Since there are only short-range AF correlations in the SC state,
the resonance intensity is roughly the product of superfluid density
and effective strength of AF correlations.
In optimally doped and underdoped samples ($x \leq x_{c}$),
the SC gap amplitude is already fixed as the pseudogap
size below $T^{*}$ and thus
the resonance energy $E_{r}$ does not shrink
towards lower energy when approaching $T_{c}$.
However, its intensity
vanishes (or becomes strongly suppressed)
when temperature is increased toward $T_{c}$, following
an order parameter-like behavior due to its proportionality to
the superfluid density.
This suggests that 
the sharp inelastic neutron resonance near optimal doping is a direct
consequence of the collective
excitations of the spin-triplet ground state.
Observed incommensurate magnetic peaks may come from a band structure effect
at the Fermi energy. There is also possibility that they are from
the formation of 
inhomogeneous stripe structure.\cite{Tranquada:1997}
It is important to note that the (energy resolution limited) collective
mode nature of the neutron scattering resonance
is best understood,
when the pairing and AF correlations are treated on {\em equal footing}.

   Recently Demler and Zhang\cite{Demler:1995} have reached a similar
result to ours in the context of SO(5) symmetry,
although details are somewhat different.
In SO(5) theory, the spin-triplet amplitude is not an order parameter, but
the generator of infinitesimal rotations between AF and SC
order parameters.
In the mixed AF+SC+spin-triplet phase that is
discussed here, the spin-triplet state acquires a nonzero order
parameter in the ground state.
The above argument which is valid near $x=x_{c}$
drastically changes with decreasing doping.
In underdoping region where the first way of transition starts
to become more operative,
the resonance energy is no longer given by twice of the maximum gap energy.
Otherwise, it would keep increasing with more underdoping, 
because the pseudogap size
increases with decreasing doping.
The intensity also does not behave like an order parameter any more,
as it does near optimal doping.
As the doping is decreased, the resonance energy becomes soft and becomes
a Goldstone mode of the AF order near half-filling.
To our opinion, there is no obvious reason for linear scaling of the
resonance energy with $T_{c}$.
At present it is uncertain whether this linear scaling is just a
coincidence or comes from a deeper theoretical origin.
It is expected that at optimal doping region or 
in slightly underdoping region
two resonance peaks can appear in inelastic neutron scattering
experiments, one from the first way of transition and the other from
the second way.

   With decreasing doping,  
the induced pairing correlations as well as             
the phase coherence-breaking AF correlations increase
in underdoped and optimally doped samples.
As a result, with decreasing doping,
the SC gap amplitude increases but superfluid density or $T_{c}$ decreases.
The resulting $(\Delta_{d})_{\mbox{max}}/K_{B}T_{c}$ ratio is strongly
doping dependent, monotonically increasing with decreasing doping
below $x_{c}$. Above $x_{c}$ (overdoping), however, it is expected
that the ratio approaches more or less the BCS mean-field value.
This feature cannot be understood in the absence of AF correlations  
which are allowed in the model.
The effective strength for the SC long-range
order is also strongly doping dependent and is largest near $x_{c}$.
It decreases below $x_{c}$ due to the increasing phase coherence-breaking
AF correlations and also above $x_{c}$ owing to the decreasing
pairing correlations, as discussed in the previous Section.
In this respect, it is not surprising to find that
the superfluid density and the SC condensation energy have their
maximum values near $x_{c}$, and decrease below
and above $x_{c}$.\cite{Tallon:2000}
In the present scenario,
the pseudogap is virtually unchanged by an applied
magnetic field, because the characteristic
energy scale for the pseudogap, $\Delta$, is much larger than
the Zeeman energy.
On the other hand,
the SC long-range order is relatively easily destroyed by it
because of its phase coherence-breaking nature.

   Finally it is worthwhile to comment on the SC condensation energy.
In the present scenario
the SC long-range order is stabilized only  
through a pair hopping process along the c-axis (due to the interlayer
coupling). It forces 
the SC condensation energy to come from the lowering of the 
c-axis kinetic energy in the SC state.
This interlayer coupling theory was already proposed by Anderson and 
others,\cite{Wheatley:1988} and many features are consistent with c-axis 
optical measurements.\cite{Basov:1994}
Recently some significant discrepancies found 
in the measured c-axis penetration 
depth in Tl$_{2}$Ba$_{2}$CuO$_{6+\delta}$ and in the prediction from the 
interlayer tunneling model, were resolved by Chakravarty 
{\it et al.}\cite{Chakravarty:1999} by subtracting fluctuation 
effects in the electronic specific heat data.
\section{gap symmetry in the SC state}
\label{section5}

   In this Section, we discuss the gap symmetry of the SC 
order parameter for cuprates.
In the previous study,\cite{Kyung:2000-2} 
for several $d$-wave type symmetries
the effective strength of pairing correlations,
$\langle \Delta^{+}_{g}(0)\Delta_{g}(0) \rangle$,
was found as 
\begin{eqnarray}
& &  \frac{n}{4N}\sum^{'}_{\vec{k}}\phi_{g}^{2}(\vec{k})
     [f(E_{-}(\vec{k}))
     +f(E_{+}(\vec{k}))]
-sign[\frac{\phi_{g} (\vec{k}+\vec{Q})}{\phi_{g}(\vec{k})}]
                                             \nonumber  \\
&\times& \frac{\Delta_{sdw}}{2UN}\sum^{'}_{\vec{k}}\phi_{g}^{2}(\vec{k})
     \frac{\Delta_{sdw}}
          {\lambda(\vec{k})}
     [f(E_{-}(\vec{k}))
     -f(E_{+}(\vec{k}))]
                           \; ,
                                                           \label{eq30}
\end{eqnarray}
where
$\lambda(\vec{k})={\sqrt{((\varepsilon(\vec{k})
                -\varepsilon(\vec{k}+\vec{Q}))/2)^{2}
                +\Delta_{sdw}^{2}}}$ and
$E_{\pm}(\vec{k})=
  (\varepsilon(\vec{k})+\varepsilon(\vec{k}+\vec{Q}))/2
   \pm \lambda(\vec{k})$.
Generally $\Delta_{g}(i)$ is defined as 
$\Delta_{g}(i)  =  \frac{1}{2}\sum_{\delta}g(\delta)
       (c_{i+\delta,\uparrow}        c_{i,\downarrow}
       -c_{i+\delta,\downarrow}      c_{i,\uparrow})$,
where $g(\delta)$ is an appropriate
gap structure factor in real space.
$\varepsilon(\vec{k})=-2t(\cos k_{x}+\cos k_{y})-\mu$ for
nearest neighbor hopping,
$\mu$ is the chemical potential controlling
the particle density $n$,
$N$ the total number of lattice sites,
$f(E)$ the Fermi-Dirac distribution function,
$\phi_{g}(\vec{k})$ the Fourier transform of $g(\delta)$, and
the summation accompanied by the prime symbol is over wave vectors
in half of the first Brillouin zone.
For local $s$-wave and extended $s$-wave symmetries, there are additional
contributions to the above equation. But the second term in Eq. ~\ref{eq30}
is still the major factor determining 
whether or not
$\langle \Delta^{+}_{g}(0)\Delta_{g}(0) \rangle >
 \langle \Delta^{+}_{g}(0)\Delta_{g}(0) \rangle_{0}$.

   Purely from a symmetry reason are ruled out   
several gap symmetries such as  
local $s$-wave ($\phi (\vec{k})=1$) and $d_{xy}$
($\phi(\vec{k})=2\sin k_{x} \sin k_{y}$), both of which have 
$\phi (\vec{k}+\vec{Q})=\phi (\vec{k})$.
In the real space representation, 
a pair with a local $s$-wave symmetry
involves 
an upspin (downspin) electron and 
a downspin (upspin) electron at the same site $i$. 
A pair with a $d_{xy}$ symmetry
the combination of
an upspin (downspin) electron in site $i$ and
the downspin (upspin) electrons in the next-nearest neighbors
with an appropriate sign.
Thus the pair configurations of the above two symmetries are directly  
against that dictated by the strong local 
Coulomb repulsion $U$.
Eventually the most stable gap symmetry  
among various possible types with 
$\phi (\vec{k}+\vec{Q})=-\phi (\vec{k})$,
is determined by the band structure 
near the Fermi surface.
Near half-filling
$|\phi (\vec{k})|$ of $d_{x^2-y^2}$ symmetry 
($\phi (\vec{k}) = \cos k_{x} - \cos k_{y}$) is much larger than,
for instance, that of extended $s$-wave 
($\phi (\vec{k}) = \cos k_{x} + \cos k_{y}$) near the Fermi surface.
It enables $d_{x^2-y^2}$ gap symmetry to be realized in most cases. 
Far away from half-filling, however,
the extended $s$-wave form factor becomes important and 
there is possibility that 
the SC gap contains a substantial fraction of 
extended $s$-wave component, leading to a
$d_{x^2-y^2}+is$ gap symmetry.

   Recently there is some 
controversy over the gap symmetry in electron-doped 
cuprates Nd$_{2-x}$Ce$_{x}$CuO$_{4-\delta}$ and 
         Pr$_{2-x}$Ce$_{x}$CuO$_{4-\delta}$,
namely, isotropic $s$-wave against anisotropic $d$-wave.\cite{Kokales:2000}
According to the above principle of finding the most stable gap symmetry,  
local (or isotropic in momentum space) $s$-wave is ruled out from the 
beginning. It is due to its incompatibility with the underlying strong 
local Coulomb repulsion.
The correct gap symmetry for electron doped cuprates
should be extended $s$-wave or 
$d_{x^2-y^2}$ or a mixture of these symmetries, depending on 
the projected density of states at the Fermi energy for a given symmetry.
The gap of an extended $s$-wave symmetry vanishes only at the AF zone 
boundary.   
According to ARPES experiments\cite{King:1993} 
and band structure calculations,\cite{Massidda:1989}
however, the Fermi surface of electron doped cuprates is
likely to be circular near the zone center.
In this situation a gap node of the extended $s$-wave symmetry does not appear  
in the Fermi surface. Consequently 
this symmetry, if it is correct, 
might have been interpreted as the isotropic $s$-wave in previous experiments. 
This feature may imply that    
the observed pairing symmetry itself does not tell us any crucial information 
about the mechanism of superconductivity.
\section{Comparison with some leading theories
         based on the Hubbard and $t-J$ models}
\label{section6}

   In this Section we compare the present approach
with some leading theories for the high-T$_{c}$ superconductivity.
Perhaps the closest theories to our approach are 
those based on the Anderson's RVB state
in one way or another. 
Anderson and his co-workers first applied a mean-field
approximation\cite{Baskaran:1987} to the $t-J$ model.
The mean-field theory studied by these authors
and by others\cite{Ruckenstein:1987,Kotliar:1988,Affleck:1988,Dombre:1989,%
Lederer:1989,Ubbens:1992}
has been a starting point for further development
of the theory such as $1/N$ expansion theory\cite{Grilli:1990}
and gauge theory\cite{Baskaran:1988,Ubbens:1994}
of the $t-J$ model.
In slave boson theory of the $t-J$ model,
typically two mean-field order parameters are considered
\begin{eqnarray}
\chi_{ij} &=& \langle f^{+}_{i,\sigma}f_{j,\sigma} \rangle  \; ,
                                             \nonumber  \\
\Delta_{ij} &=& \langle f_{j,\uparrow}    f_{i,\downarrow}
                    -f_{j,\downarrow}  f_{i,\uparrow} \rangle
                                                      \; ,
                                                         \label{eq40}
\end{eqnarray}
together with $\langle b_{i} \rangle$
In the slave boson representation,
a physical electron is decomposed
into a spinon (fermion) and a holon (boson),
$c^{+}_{i,\sigma} = f^{+}_{i,\sigma}b_{i}$.

   Depending on the vanishing or nonvanishing of $\Delta_{ij}$ and
$\langle b_{i} \rangle$, the doping and temperature plane is divided
into four regions.\cite{Ubbens:1994}
Region I with $\Delta_{ij} = 0$ and $\langle b_{i} \rangle \ne 0$ is
a Fermi liquid phase.
Region II with $\Delta_{ij} \ne 0$ and $\langle b_{i} \rangle = 0$ is
the spingap phase, in which a $d$-wave gap appears in the fermion
spectrum without Bose condensation of holons.
Region III with $\Delta_{ij} \ne 0$ and $\langle b_{i} \rangle  \ne 0$
indicates SC long-range order in physical electrons.
Region IV with $\Delta_{ij} = 0$ and $\langle b_{i} \rangle  = 0$
is designated as the strange metal phase, because it shows
various non-Fermi liquid features.

   In many respects, the slave boson mean-field
theory of the $t-J$ model\cite{Baskaran:1987,Ruckenstein:1987,Kotliar:1988,%
Affleck:1988,Dombre:1989,Lederer:1989,Ubbens:1992}
has shed some important insight into the microscopic understanding
of the cuprate superconductors.
This is because
the predicted phase diagram is, at least, qualitatively
consistent with experiments, and
the pseudogap is closely related to a spingap,
and furthermore
it starts from the microscopic model
as opposed to other
phenomenological models.
However, there are also some serious problems
with the slave boson mean-field theory,
as noted by Ubbens and Lee.\cite{Ubbens:1994}

   One of them
is that the temperature scale for Bose condensation of holons
is too high. Furthermore the maximum $T_{c}$, which is determined
by the two lines
$\Delta_{ij} \ne 0$ and $\langle b_{i} \rangle  \ne 0$, occurs
at too small doping concentration ($x < 0.06$).
At this doping level,
the SC long-range order even does not appear in cuprate superconductors.
Close to half-filling, several exotic phases have been reported to
be stable such as mixed phases\cite{Kotliar:1988}
(equivalently $\pi$-flux phases\cite{Affleck:1988}),
dimerized phases,\cite{Affleck:1988,Dombre:1989} and
staggered flux phases.\cite{Lederer:1989,Ubbens:1992}
It is unclear whether these states are realized or not
in cuprates.
In a recent paper\cite{Kyung:2000-3} we argued that
these problems can be naturally
resolved, when AF correlations, the weaknesses of a mean-field approximation,
and the limitation of the $t-J$ model near half-filling
are properly taken into account.

   In the original RVB theory,\cite{Anderson:1987} Anderson suggested that 
the RVB ground state may be obtained by Gutzwiller 
projection on the BCS ground state.
As shown in the previous study,\cite{Kyung:2000-2} however,
strong local Coulomb repulsion $U$ induces the AF and pairing correlations
{\em at the same time}.
This makes  
one Gutzwiller projection on the noninteracting electrons enough to 
yield the RVB ground state.
Unless the interlayer coupling is introduced in the RVB based theories,
they suffer from having only one energy scale $\Delta$ which has to explain 
$T^{*}$ and $T_{c}$ at the same time.
Some studies on the $t-J$ model at half-filling
or the Heisenberg model
showed that a mixed phase\cite{Kotliar:1988} 
(equivalently a $\pi$-flux phases\cite{Affleck:1988})
is the ground state of the model.
However, the observed energy dispersion
in the insulating cuprates
Sr$_{2}$CuO$_{2}$Cl$_{2}$\cite{Wells:1995}
and Ca$_{2}$CuO$_{2}$Cl$_{2}$,\cite{Ronning:1998}
is different from what the flux phase 
predicts. The observed band dispersions
along $(\pi/2,\pi/2)-(\pi,0)$ and
$(\pi/2,\pi/2)-(0,0)$ are not identical and 
the energy dispersion appears quadratic near 
$(\pi/2,\pi/2)$ point instead of linear.
In fact the experiments are more consistent with our calculations based 
on the Hubbard Hamiltonian\cite{Kyung:2000-2}.
Although these differences (the Hubbard vs. $t-J$ Hamiltonians) are rather 
quantitative, it is not clear whether in some subtle issues associated with
our conjectures these Hamiltonians give the same answer or not.

   The mechanism to superconductivity proposed in our previous and this papers 
is fundamentally different from the conventional one in which
superconductivity is driven by exchange of 
some bosonic degrees of freedom such as phonons,
spin and charge fluctuations and so on.
It suggests that the exchange of spin fluctuations\cite{Bickers:1989} 
cannot be
the leading mechanism to high temperature superconductivity, at least,
in the Hubbard type model.
This is because this approach completely neglects the leading pairing 
interaction directly induced from the local Coulomb repulsion. 
But instead it depends on the residual pairing interaction from the 
AF correlations which were already induced 
(simultaneously with the pairing correlations)
from the Coulomb repulsion.
\section{Numerical studies for the Hubbard and $t-J$ models}
\label{section7}

   There have been several numerical studies on the Hubbard and 
$t-J$ models in two dimensions.\cite{Dagotto:1994}
Among many important issues,
the existence of pairing correlation which may lead to a SC 
instability at low temperature is of particular 
importance.
Several methods including exact diagonalization (ED), quantum Monte
Carlo (QMC) 
and density matrix renormalization group (DMRG) 
support with decreasing temperature 
growing $d$-wave pairing correlations away from 
half-filling.
This is consistent with a weak coupling RG study of the 
Hubbard model.\cite{Schulz:1987}
We believe what many numerical methods such as 
ED, QMC and DMRG
find is the pairing correlations induced from the local 
Coulomb repulsion in the presence 
of the induced AF correlations.
However, the fact that in the numerical calculations
the pairing strength increases with decreasing doping 
just as the AF correlations do, might have given a 
misleading impression that 
the pairing correlations come from the exchange of 
spin fluctuations.

   On the other hand, 
a constrained path quantum Monte Carlo 
(CPQMC) method\cite{Zhang:1997} appears to show  
a negative answer to the existence 
of pairing correlations in the Hubbard model.
This result is in direct conflict with 
that of the RG study.\cite{Schulz:1987}
Although it is valid in the weak coupling limit,
the RG result cannot be qualitatively wrong. 
It is because of its systematic nature and 
the general experience that many weakly interacting
systems are more or less smoothly connected to corresponding
strong coupling systems.
Thus in the CPQMC method, there is possibility that 
away from half-filling their ill-suited trial 
wave function (SDW solution) 
generates
mainly the states with AF correlations but neglects those with pairing
correlations. 
Or in the process of constraining paths to avoid
the fermion sign problem,
the $d$-wave pairing correlations are strongly suppressed.
Indeed Zhang's calculations may suggest the fate of those theories based on
purely magnetic correlations such as the exchange of spin fluctuations,
magnetic polarons and so on.
\section{Hole-doped cuprates and heavy-fermion superconductors}
\label{section8}

   The present scenario may also apply 
without any drastic modification
to electron-doped high-$T_{c}$
superconductors as well as heavy-fermion superconductors.
The main qualitative difference between hole-doped cuprates and 
the above compounds\cite{Comment20}
is that for the latter local magnetic moments from 
Ce or U atoms residing between conducting planes
play an important role in determining the phase diagram 
and electronic properties.
Electrons in the CuO$_{2}$ plane interact with 
local magnetic moments from Ce or U atoms. 
This enables the AF correlations in the CuO$_{2}$ plane to easily induce 
AF correlations in those atoms, enhancing the overall 
AF fluctuations in the CuO$_{2}$ plane.
Then the close balance between the pairing and AF
correlations is broken in favor of the latter
in the conducting electrons. 
As a result, AF long-range phase significantly increases, 
and accordingly SC phase shrinks and remains at the edge of the 
AF phase boundary.
Because of the resulting imbalance of the two correlations 
in the conducting plane,
there is 
possibility that the pseudogap state is completely 
destroyed or remains at most in a narrow region of the $x-T$ plane.
The corresponding SC gap symmetry is determined  
by the relative sign of 
$\phi_{g} (\vec{k}+\vec{Q})$ vs. $\phi_{g}(\vec{k})$ and 
the shape of the 
Fermi surface. 
In general it can be other than a $d$-wave symmetry.
Thus, high temperature superconductors for both hole-doped and 
electron-doped cuprates as well as heavy-fermion superconductors may be  
understood in the 
{\em unified} framework within the present scenario.
We also argue that differences observed in hole-doped and electron-doped
cuprates do not come from the different signs of charge carriers.
They are from the presence of magnetically active atoms (Ce) between
CuO$_{2}$ planes in the latter.
If this turns out to be correct, the correct model Hamiltonian 
for the heavy-fermion superconductors would be 
the periodic Anderson model or the Kondo lattice model with 
{\em strongly correlated} conducting electrons  
instead of with the noninteracting electrons.
Various studies of the superconductivity  
on the basis of spin fluctuations
in these compounds and even 
in organic superconductors 
should be reconsidered for their validity. 
\section{Maximizing the SC critical temperature $T_{c}$}
\label{section9}

    It is also interesting to
discuss the issue of maximizing the 
SC critical temperature $T_{c}$ on the basis of the present scenario.
There are roughly three ways of increasing $T_{c}$.
First, $T_{c}$ can be increased by a stronger interlayer coupling.
In the family of Bi$_{2}$Sr$_{2}$Ca$_{n-1}$Cu$_{n}$O$_{2n+4}$ (BSCCO), 
TlBa$_{2}$Ca$_{n-1}$Cu$_{n}$O$_{2n+3}$ (T1BCCO), 
Tl$_{2}$Ba$_{2}$Ca$_{n-1}$Cu$_{n}$O$_{2n+4}$ (T2BCCO), and 
HgBa$_{2}$Ca$_{n-1}$Cu$_{n}$O$_{2n+2}$ (HBCCO) where $n=$1, 2, 3 and 4,
$T_{c}$ for larger n is higher than that of smaller n.
Except for T1BCCO, however, all the others show their highest 
$T_{c}$'s at $n=3$. It indicates the saturation of the interlayer coupling
strength with increasing the number of layers in a unit cell.
Making the interlayer region more metallic is also suggestive to 
enhance the interlayer coupling.
Second, $T_{c}$ can be increased by decreasing the local Coulomb repulsion $U$.
When HBCCO compound is subjected to a high pressure of 30 GPa, 
$T_{c}$ increases from 133 K to 164 K.
Applying pressure causes the increase of hopping integral between nearest 
neighbors, effectively decreasing the local Coulomb repulsion $U$.
Since Coulomb repulsion is also a driving force to the pairing correlations,
$U$ cannot be arbitrarily small to achieve the highest $T_{c}$.
There must be an optimal strength $U$ (presumably $U \sim W$) 
in which $T_{c}$ is maximized.
Third, $T_{c}$ can be also increased by breaking the close balance between 
the pairing and AF correlations in favor of the former.
In principle this can be achieved by putting some atoms 
between the CuO$_{2}$ planes, which respond {\em diamagnetically} to
the phase coherence-breaking AF fluctuations in such a way that the overall 
strength of AF correlations are reduced in the CuO$_{2}$ planes.
It is exactly opposite to what happens in electron-doped cuprates.
Although the implementation of this possibility is uncertain 
in a real situation, this last 
way can be most effective to maximize $T_{c}$.
\section{Conclusion}
\label{section10}

   Before closing we comment on a few points. Since the 
conclusion of 
the present paper crucially depends on the three conjectures,
it is important to prove or disprove them,
of course, on the basis of 
fully systematic or rigorous calculations.
It is believed that the mechanism to superconductivity and the interplay
between antiferromagnetism and superconductivity expounded in this paper
can be applicable to heavy-fermion superconductors as well as
organic superconductors.
A stripe issue has not been specifically addressed in this paper.
Through the microscopic separation of hole-rich regions
from antiferromagnetically correlated regions,\cite{Zaanen:1989}
the stripe structure tends to maintain the AF correlations more effectively
than the other case in which
they are uniformly suppressed by doped holes.
Then it is not difficult to expect that in the stripe state
$T_{c}$ and $T^{*}$ are suppressed by some amount
from those values in the uniform state.
We believe that the differences in the stripe state
and in the uniform state
are mainly quantitative. 
The conclusion reached in this paper
is not expected to qualitatively change in the presence of stripe structure.
However, stripe-type structure can be important in the low doping region
to stabilize both pairing and AF
correlations at the same time.\cite{Kyung:2000-3}

   In some of magnetic experiments such as NMR,\cite{Reyes:1991}
another crossover temperature
$T^{0}$ (larger than $T^{*}$) is often identified,
at which Knight shift shows its maximum.
This feature can be easily understood on the basis of the competing
nature of pairing and AF correlations as well as
of the phase diagram obtained in Ref. ~\cite{Kyung:2000-3}
The obtained mean-field phase diagram shows that the mean-field AF phase 
line stays above the mean-field SC phase line for $x \leq x_{c}$.
As was discussed in Section~\ref{section3},
these AF and SC mean-field phase lines should be interpreted
as the onset ($T^{0}$)
of short-range AF correlations 
and as the onset ($T^{*}$) of short-range pairing correlations
(pseudogap), respectively.
For $T^{*} < T < T^{0}$, the two correlations compete, while they
grow with decreasing temperature.
In spin-lattice relaxation rate which picks up strongly
the $\vec{q}=\vec{Q}$ component, for $T^{*} < T < T^{0}$
the contribution from the AF correlations dominates that from
the pairing correlations so that $1/T_{1}T$ keeps increasing until
it starts to decrease at $T^{*}$.
On the other hand,
Knight shift which picks up
the $\vec{q}=0$ component and thus is unaware of the growing AF
correlations, is more strongly influenced by the increasing
pairing correlations.
Consequently Knight shift reaches its maximum at $T^{0}$, and starts to
slowly decrease below it and then rapidly decrease below $T^{*}$.

    Based on experimental results and our previous theoretical work, 
we have conjectured 
a microscopic theory of high temperature superconductivity.
In this conjecture,
SC and AF long-range orders are driven by 
interlayer coupling.
Strictly in two dimensions, the microscopic Hubbard model has an 
RVB insulator-to-metal transition 
at $x=x_{c}$ near optimal doping 
for zero temperature, leading to a QCP, 
and one of the crossover lines is given by the pseudogap 
temperature $T^{*}$.
We argued that various singular and non-Fermi liquid properties observed  
near optimal doping are due to the presence of this QCP.
In our conjecture,
the crossover line $T^{*}$ also practically divides the SC  
region into two,
depending on the doping level with respect to $x_{c}$.
For $x \leq x_{c}$ the SC state has significant AF correlations, while 
for $x > x_{c}$ it has virtually no AF correlations, thus 
justifying the conventional 
BCS theory based on the noninteracting electrons.
Inelastic neutron scattering resonance and systematically reduced 
superfluid density in the SC state below $x_{c}$ 
have their natural explanations in the 
present scenario.
The present approach supports
interlayer pair tunneling model in which
the SC condensation energy comes from the lowering of the 
c-axis kinetic energy in the SC state.
Comparison of the present scenario with some of the leading theories 
based on the Hubbard and $t-J$ models was given. 
The generic features of
both hole-doped and electron-doped cuprates as well as 
heavy-fermion superconductors may be understood in the 
{\em unified} framework within the present picture.
\acknowledgements

   The author would like to thank A. M. Tremblay for numerous help and 
discussions throughout the work. 
Without his continuous encouragement and support, this work would have been
impossible.
The author also thanks C. Bourbonnais, N. Dupuis, P. Fournier, 
A. J. Millis, S. H. Pan, D. S\'{e}n\'{e}chal and 
all the participants of the Canadian Institute for Advanced Research
(CIAR) Meetings,
for stimulating discussions.
The present work was supported by a grant from the Natural Sciences and
Engineering Research Council (NSERC) of Canada and the Fonds pour la
formation de Chercheurs et d'Aide \`a la Recherche (FCAR) of the Qu\'ebec
government.
%
%

%
%
%
%
\begin{figure}
 \vbox to 13.0cm {\vss\hbox to -5.0cm
 {\hss\
       {\includegraphics{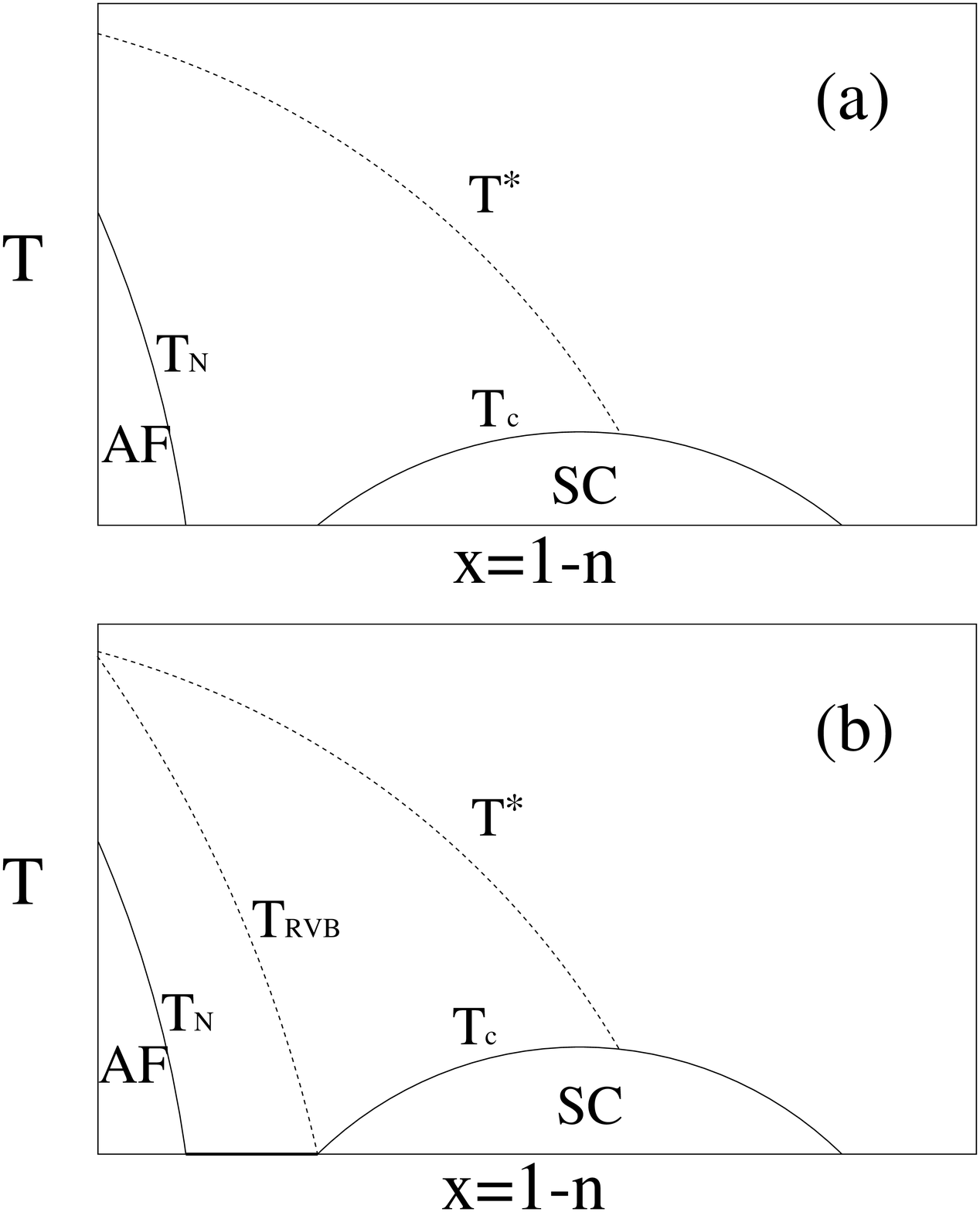}
       }
  \hss}
 }
\caption{Schematic phase diagrams in doping ($x=1-n$)
         and temperature ($T$) plane based on (a) experiments and 
         (b) our previous work, in the presence of the interlayer 
         coupling. $T_{c}$ and $T_{N}$ denote the 
         SC and AF transition temperatures, respectively. $T^{*}$
         is the pseudogap temperature, while $T_{RVB}$ in (b) is 
         a temperature where the AF correlations become strong.
         The thick solid line in (b) denotes the insulating 
         RVB ground state with a spingap.} 
\label{fig1}
\end{figure}
%
%
%
%
\begin{figure}
 \vbox to 15.0cm {\vss\hbox to -5.0cm
 {\hss\
       {\includegraphics{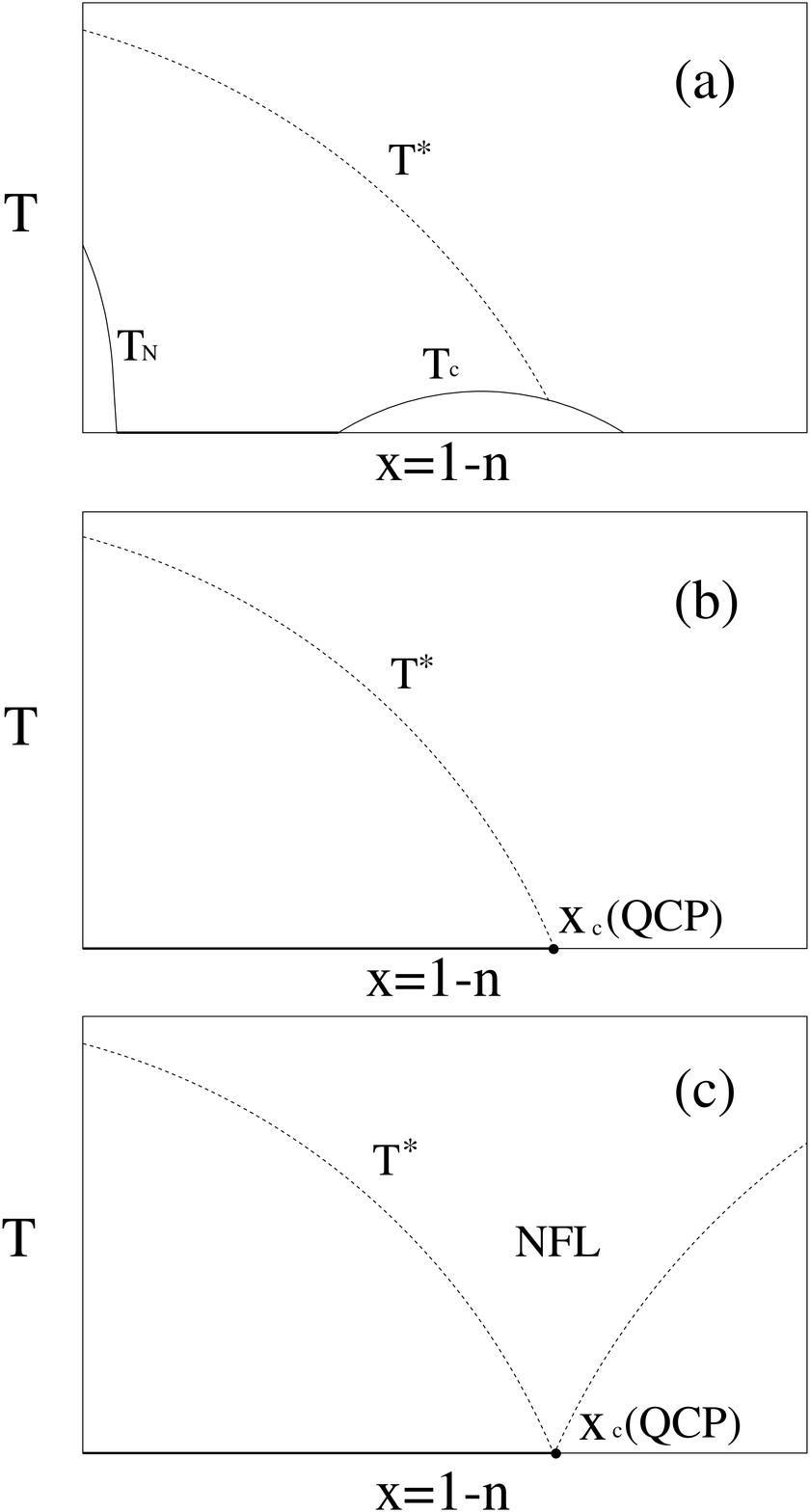}
       }
  \hss}
 }
\caption{Schematic phase diagrams in doping ($x=1-n$)
         and temperature ($T$) plane (a) with a reduced 
         interlayer coupling, and (b) and (c) without it.
         $T_{c}$, $T_{N}$, and $T^{*}$ are 
         the same as in Fig. ~\ref{fig1}.
         The thick solid lines denote the insulating RVB ground state with 
         a spingap $\Delta$, and without the interlayer coupling
         these lines terminate at $x=x_{c}$
         for $T=0$ shown as a black dot in (b) and (c).}
\label{fig2}
\end{figure}

\begin{references}
\bibitem{Comment6.5}    
The readers should not be confused with SC long-range order
$\langle \Delta_{g}(0) \rangle \ne 0$ and with pairing correlations
$\langle \Delta^{+}_{g}(0)\Delta_{g}(0) \rangle >
 \langle \Delta^{+}_{g}(0)\Delta_{g}(0) \rangle_{0}$.
$\langle \cdots \rangle_{0}$ means an expectation value evaluated
in the ground state of noninteracting electrons.
$\Delta_{g}(i)$ is a pair operator with an appropriate gap symmetry,
which is defined in Section~\ref{section5}.
\bibitem{Comment9} In Fig. ~\ref{fig1}(b), $T_{RVB}$ at half-filling is  
   schematically drawn to be close to $T^{*}$. In general
   $T_{RVB}$ can be smaller or larger than
   $T^{*}$ at $n=1$.
   However, the argument remains the same regardless of its location.
\bibitem{Comment10} In the previous study we did not realize that the  
                    local Coulomb repulsion can also induce the $d$-wave 
                    pairing correlations. But the conclusion equally applies
                    to the present problem.
\bibitem{Comment20} For heavy-fermion compounds the relevant energy 
                   scale appears much smaller than that of hole-doped 
                   and electron-doped cuprates because of much 
                   lower $T_{N}$ and $T_{c}$.
\end{references}
\end{document}